\documentclass[twoside,12pt]{article}
\usepackage{amsfonts}
\usepackage{amssymb}
\usepackage[dvips]{graphicx}
\usepackage{latexsym}
 \newcommand{\overskrift}[1]{\vspace{6.0mm}\noindent\textbf{#1}\vspace{1.5mm}}

 \newcommand{\beq}{\begin{equation}}
 \newcommand{\eeq}{\end{equation}}
 \newcommand{\bea}{\begin{eqnarray}}
 \newcommand{\eea}{\end{eqnarray}}
 \newcommand{\lpart}{\raise.3ex\hbox{$\stackrel{\leftarrow}{\partial}$}}
 \newcommand{\rpart}{\raise.3ex\hbox{$\stackrel{\rightarrow}{\partial}$}}
 \newcommand{\ldr}{\raise.3ex\hbox{$\stackrel{\leftarrow}{\delta^r}$}}

\begin{document}
\begin{flushright}

hep-ph/0507315

\end{flushright}

\begin{center}
\vspace{3.0cm}

{\bf \Large Suppressing Super-Horizon Curvature Perturbations?}

\vspace{1.0cm}

{\bf Martin S. Sloth\footnote{\texttt{e-mail:
sloth@physics.ucdavis.edu}}}

\vspace{0.5cm}

 \emph{University of California, Davis}

 \emph{Department of Physics}

 \emph{CA 95616, USA}

\end{center}
\vspace{0.5cm} {\abstract{\noindent We consider the possibility of
suppressing superhorizon curvature perturbations after the end of
the ordinary slow-roll inflationary stage. This is the opposite of
the curvaton limit. We assume that large curvature perturbations
are created by the inflaton and investigate if they can be diluted
or suppressed by a second very homogeneous field which starts to
dominate the energy density of the universe shortly after the end
of inflation. We show explicit that the gravitational sourcing of
inhomogeneities from the more inhomogeneous fluid to the more
homogeneous fluid makes the suppression difficult if not
impossible to achieve.}}

\vspace{1.0cm}

\newpage
\section{Introduction}

The inflationary scenario is now the by far most well established
scenario for generating the observed microwave background
radiation (CMB) anisotropies and in addition it explains the
flatness and isotropy of the universe \cite{inflation}.

The curvaton mechanism was invented to show how scenarios
otherwise ruled out by observations could in fact still be viable
inflationary candidates\footnote{Related ideas regarding the creation
  of adiabatic density perturbations from initial isocurvature perturbations was discussed
  already in \cite{Mollerach:1989hu,Linde:1996gt}.} \cite{Enqvist:2001zp,Lyth:2001nq,Moroi:2001ct}. Here we pursue a similar
philosophy but in the
opposite direction. The constraint from the power of tensor modes
generated during inflation is one of the most basic and rigid
constraints on inflation as it confines the inflationary energy
scale to or below the Grand Unified Theory (GUT) scale. Also to
match the observed level of adiabatic scalar metric perturbations
at the level of $10^{-5}$, we have to face some great fine-tuning
problems. Here we want to address if there is any circumstance's
under which the second of these constraints can be softened. We
find that it is important to understand how strong our
experimental constraints on the inflationary dynamics are and how
sensitive they are to the assumptions we make. We find that it is
very difficult if not impossible to circumvent these constraints
and the upper bounds are indeed very robust.

The philosophy driving us is very simple. We consider a very
homogeneous fluid, sub-dominant during inflation, which comes to
dominate the energy density of the universe only a few e-foldings
after the end of inflation. As we are in this way pumping the
universe with a homogeneous fluid we might expect the wrinkles in the
original radiation fluid left over by the decay of the inflaton to
be suppressed relatively to the total energy density of the
universe. This is the exact opposite limit of the curvaton, where
the radiation fluid was initially thought to be very homogeneous
and wrinkles in the overall density was created by the less
homogeneous curvaton fluid which subsequently came to dominate the
energy density. However, as we will see there is a serious
complication in this new limit as compared to the curvaton limit.
While in the curvaton scenario one could consistently ignore the
effect of the gravitational potential as it initially vanishes, in
the new scenario the gravitational potential is non-zero and can
not be ignored. In fact it sources gravitationally inhomogeneities
from the inhomogeneous radiation fluid to the second more
homogeneous fluid which will in general not stay homogeneous for
very long after the end of inflation. This is the reason why in
the case where the potential of the second field is a simple mass
term and its decay rate has trivial time-dependence, the effect is
less than order one.

We go on to investigate the example with the massive field
replaced with an axion field. The periodic nature of the axion
potential suppresses entropy perturbations in the axion fluid and
it appears to avoid some of the problems with the massive fields.
However, as we discuss, this example suffers very similar
problems.

\section{Suppressing Super-Horizon Scalar Perturbations?}

Let us consider how one might think that large adiabatic density
fluctuations created by the inflaton can be suppressed during a
post inflationary era. We will assume that the inflaton or its
decay product dominates just after the end of inflation and as the
inflaton decays into radiation the curvature fluctuations
generated during inflation will be inherited by the radiation
fluid. Consider for example a light axion field ($\sigma$) frozen
during inflation. Due to its periodic potential it contributes
with no or vanishing density fluctuations. In the post
inflationary era it oscillates in its potential and decays into
radiation while it slowly starts to dominate over the inflaton or
the inflaton decay products. In this way it appears that it will dilute the
curvature perturbations generated by the inflaton.

One might at first think of it naively in the following way.
Normally in the simple single field case where the inflaton decays
into radiation\footnote{Recently, a similar argument appeared in a
  pre-print by Bartolo, Kolb and Riotto \cite{bkr}. In
the first version they did not consider the gravitational sourcing
of inhomogeneities from the more inhomogeneous fluid to the more
homogeneous one and as we show below this poses a problem for the
mechanism.}
 \beq \label{drh}
\frac{\delta\rho}{\rho}=\frac{\delta\rho_r}{\rho_r}\simeq const~.
 \eeq
In the case of the axion eq.(\ref{drh}) is still valid at the end
of inflation if $\delta\rho_a/\rho_a\simeq 0$ and
$\rho_a<<\rho_r$. Here $r$ denotes radiation and $a$ denotes the
axion. Later when the axion oscillates in its potential and behave
like matter we find $\rho_a=\rho_m>>\rho_r$, while $\delta\rho_a$
can still be neglected. At this point
 \beq\label{drh1}
\frac{\delta\rho}{\rho}=\frac{\delta\rho_r}{\rho_m}=\frac{\rho_r}{\rho_m}
\frac{\delta\rho_r}{\rho_r} ~.
 \eeq
Thus the density perturbations is suppressed by the value of
$\rho_r/\rho_m$ as the axion decays into radiation. We are pumping
homogeneous matter into the system, diluting away the original
density perturbations.

One can also note that the change in the superhorizon curvature
perturbation $\dot\zeta$ is proportional to the entropy
perturbation $S_{ra}=3(\zeta_r-\zeta_a)$. So while in the curvaton
limit the curvature perturbation in the radiation is subdominant
$\zeta_r<<\zeta_a$ in opposite limit we have $\zeta_r>>\zeta_a$
and thus $\dot\zeta$ has the opposite sign and $\zeta$ decreases
instead of increasing.

Even if this picture captures some of the right physics it is not
fully correct since it ignores the gravitational coupling between
the two fluids which will source density perturbations from the
more inhomogeneous fluid to the more homogeneous one. To account
for this in detail we need to consider a fully consistent
treatment of the density perturbations and calculate the gauge
invariant adiabatic scalar curvature perturbations.

As we mentioned in the introduction, the scenario we consider here
is analogous to the by now familiar curvaton scenario. The
inflationary era is ended by the decay of the inflaton into
radiation and the $\sigma$-field, a massive scalar field
subdominant and frozen during inflation, subsequently starts to
oscillate in its potential. At this point it behaves like
non-relativistic dust and it soon starts to dominate the energy
density of the universe. As the $\sigma$-field later decays into
radiation the entropy perturbations in the $\sigma$-field fluid
vanishes and its eventual density perturbations have been
converted into adiabatic curvature perturbations in the radiation
fluid.

In the most simple case, the potential of the initially light
$\sigma$-field can be assumed to be just the quadratic one, such
that the Lagrangian simply becomes
 \beq
\mathcal{L}_{\sigma} =
\frac{1}{2}\dot\sigma^2-\frac{1}{2}(\nabla\sigma)^2-\frac{1}{2}m^2\sigma^2~.
 \eeq
and the background equation of motion yields
 \beq \label{backg1}
\ddot\sigma+3H\dot\sigma+m^2\sigma = 0~,
 \eeq
where the Hubble parameter $H$ is determined by the Friedmann
equation
 \beq \label{backg2}
H^2
=\frac{1}{3}\left(\rho_r+\frac{1}{2}\dot\sigma^2+\frac{1}{2}m^2\sigma^2\right)
 \eeq
and the continuity equation for the radiation fluid energy density
$\rho_r$
 \beq \label{backg3}
\dot\rho_r+4H\rho_r=0~.
 \eeq

Now let us consider the equations governing the perturbations in
the longitudinal gauge where the perturbed metric is
 \beq
ds^2 = -(1+2\Phi)dt^2 + a^2(t)(1-2\Psi)\delta_{ij}dx^idx^j~,
 \eeq
and we can take $\Phi=\Psi$ in the absence of anisotropic stress.
On superhorizon scales we can neglect $k/a$ terms and one finds by
completely standard arguments the following system of equations
for the perturbations \cite{Mukhanov:1990me}
 \bea \label{pert1}
-3H(H\Phi+\dot\Phi)&=&
\frac{1}{2}(\dot\sigma\dot{\delta\sigma}+m^2\sigma\delta\sigma
-\dot\sigma\Phi+\rho_r\delta_r)~,\\ \label{pert2}
\dot\delta_r-4\dot\Phi&=&0~,\\ \label{pert3}
\ddot{\delta\sigma}+3H\dot{\delta\sigma}+m^2\delta\sigma&=&4\dot\sigma\dot\Phi-2m^2\sigma\Phi~,
 \eea
where $\delta_r\equiv \delta\rho_r/\rho_r$ and $\delta\rho_r$,
$\delta\sigma$ are the perturbations in the radiation energy
density and of the $\sigma$-field respectively. It is convenient
to note also that the energy density perturbation
$\delta\rho_{\sigma}$ and the pressure perturbation $\delta
p_{\sigma}$ in the $\sigma$-field are
 \bea \label{deltarho}
\delta\rho_{\sigma}&=&
\dot\sigma\dot{\delta\sigma}+m^2\sigma\delta\sigma
-\dot\sigma^2\Phi~,\\
\label{deltap} \delta p_{\sigma}&=&
\dot\sigma\dot{\delta\sigma}-m^2\sigma\delta\sigma
-\dot\sigma^2\Phi~.
 \eea

We are interested in the evolution of the curvature perturbation
after the end of inflation when the universe is initially
dominated by a radiation fluid such that $H=1/(2t)$. There is a
number of papers in the literature dealing specifically with mixed curvaton and
inflaton perturbations, some of those are \cite{Enqvist:2001zp,Lyth:2001nq,
Moroi:2001ct,Lyth:2002my,Sloth:2002xn,Malik:2002jb,Gordon:2002gv,Lyth:2003ip,Gordon:2003hw,Gupta:2003jc,Langlois:2004nn,Ferrer:2004nv,Lazarides:2004we,Malik:2004tf}.
We find it useful to review the results of \cite{Langlois:2004nn}
below as we will compare to the results obtained there in order to
fully illustrate the difficulty of the superhorizon suppression of
curvature perturbations. In \cite{Langlois:2004nn}, the solution
for the background $\sigma$-field in the radiation dominated
regime was written in the form
 \beq
\sigma = \sigma_*A\frac{J_{1/4}(mt)}{(mt)^{1/4}}\qquad,\qquad
A\equiv \frac{\pi}{2^{1/4}\Gamma(3/4)}~,
 \eeq
and the solution to the perturbation equation was given by
 \bea \label{dsigma1}
\delta\sigma &\simeq&
\frac{A}{(mt)^{1/4}}\left[\left(\delta\sigma_*-\frac{1}{2}\Phi_*\sigma_*\right)
J_{1/4}(mt)+\Phi_*\sigma_*mtJ_{-3/4}(mt)\right]\nonumber\\
\label{dsigma2}
&=&\frac{\delta\sigma_*}{\sigma_*}\sigma+t\dot\sigma\Phi_*\nonumber~.
 \eea
Above, the initial conditions for the perturbation at the
beginning of the first radiation dominated era have been defined
by $\delta\sigma=\delta\sigma_*$, $\dot{\delta\sigma}=0$,
$\Phi=\Phi_*$, $\dot\Phi=0$ and from the perturbation equations
follows that initially we also have $\delta_r\simeq -2\Phi_*$.

It then follows that \cite{Langlois:2004nn}
 \beq\label{drho2}
\frac{\delta\rho_{\sigma}}{\rho_{\sigma}}\simeq
2\frac{\delta\sigma_*}{\sigma_*}-\frac{3}{2}\frac{\dot\sigma^2}{\rho_{\sigma}}\Phi~,
 \eeq
from which we can calculate the change in the superhorizon
curvature perturbation.

The change of the superhorizon curvature perturbation is given by
the non-adiabatic pressure perturbation
 \beq
\dot\zeta = -\frac{H}{\rho+p}\delta P_{nad}
 \eeq
where the non-adiabatic pressure perturbation between two fluids
denoted by subscript $1$ and $2$ is
 \beq
\delta P_{nad} =
\frac{(c_1^2-c_2^2)(1+c_1^2)(1+c_2^2)\rho_1\rho_2}{(1+c_1^2)\rho_1+(1+c_2^2)\rho_2}S_{12}
 \eeq
and the entropy perturbation $S_{12}$ between the two fluids is
 \beq
S_{12}=3\left(\zeta_1-\zeta_2\right)~.
 \eeq
The super horizon curvature perturbation on uniform density
hypersurfaces in some given species $i$ is defined as
 \beq
\zeta_i =-\Phi
-H\frac{\delta\rho_i}{\dot\rho_i}=-\Phi+\frac{\delta_i}{3(1+w_i)}~,
 \eeq
while its sound speed $c_i$ and equation of state parameter $w_i$
are defined as
 \beq
c_i^2=\frac{\dot p_i}{\dot\rho_i}~,\qquad p_i=w_i\rho_i~.
 \eeq
Thus one has, using eq. (\ref{drho2})
 \beq \label{S2}
S_{r\sigma}=-2\frac{\rho_{\sigma}}{\dot\sigma^2}
\frac{\delta\sigma_*}{\sigma_*}~.
 \eeq
This is the result already explored in \cite{Langlois:2004nn} and
elsewhere in the curvaton literature.
It is especially interesting to note that in this case
$S_{r\sigma}$ always has the same sign irrespectively of the
initial condition for the magnitude of $\delta\sigma_*$. This is
because that even if we take $\delta\sigma_*=0$ the curvature
perturbations in the $\sigma$-field will be sourced by the
gravitational potential and grow similar to the radiation density
perturbations. Thus, unlike what one might naively expect, one
cannot obtain any suppression from a second massive field even in
the limit where its fluctuations are suppressed during the
ordinary slow-roll inflationary phase $\delta\sigma_*=0$.

One can see this in more details by evaluating explicitly the
final total superhorizon curvature perturbation $\zeta$. During an
era of $w_{total}=const$ one has
 \beq \label{zphi}
\zeta = -\frac{5+3w}{3(1+w)}\Phi~,
 \eeq
where $\zeta$ is conserved on superhorizon scales for adiabatic
perturbations and vanishing intrinsic non-adiabatic pressure
perturbations. Thus as soon as the radiation fluid and its
perturbations are completely washed away and subdominant we expect
$\zeta=const$.

To estimate the final curvature $\zeta_f$ one can again compare to
the usual curvaton case. The curvaton is taken to be a
non-relativistic dust-like fluid with density contrast $\delta_m$.
It is assumed to be decoupled from the radiation and thus the
curvature perturbations $\zeta_m$ and $\zeta_r$ are separately
conserved. The total curvature perturbation between two fluids is
given by
 \beq
\zeta =
\frac{(1+c_1^2)\rho_1\zeta_1+(1+c_2^2)\rho_2\zeta_2}{(1+c_1^2)\rho_1+(1+c_2^2)\rho_2}~,
 \eeq
so in the ordinary curvaton scenario where the cosmic fluid
consists of radiation and matter, the total curvature perturbation
becomes \cite{Wands:2000dp}
 \beq
\zeta = \frac{3\rho_m\zeta_m+4\rho_r\zeta_r}{3\rho_m+4\rho_r}~.
 \eeq

In the curvaton scenario one assumes no initial curvature
perturbation $\Phi_*=0$ and the assumed decoupling of the two
fluids is a good approximation. Even the gravitational interaction
between the two fluids can be ignored since the gravitational
potential vanishes. The assumption $\Phi_*=0$ implies $\zeta_r=0$
and vice versa, so
 \beq
\zeta_f = \zeta_m = \frac{1}{3}\delta_m
 \eeq
in this case. Using further the relation in eq.(\ref{zphi}) with
$w=0$ one finds
 \beq
\Phi_f = -\frac{3}{5}\zeta = -\frac{1}{5}\delta_m~.
 \eeq

In the more general case of mixed curvaton and inflaton
fluctuations one can also follow standard arguments and write
$\zeta_r=-\Phi_*+\frac{1}{4}\delta_r=-\frac{3}{2}\Phi_*$ which
yields \cite{Langlois:2004nn}
 \beq
\Phi_f=\frac{9}{10}\Phi_*+\frac{1}{5}S_{r\sigma}~.
 \eeq

Finally, from eq. (\ref{S2}) one obtains \cite{Langlois:2004nn}
 \beq \label{Phibotox1}
\Phi_f =
\frac{9}{10}\Phi_*-\frac{2}{5}\frac{\delta\sigma_*}{\sigma_*}~.
 \eeq
and no suppression can take place compared to the single field
case where the dynamics can be fully described by the inflaton and
its decay products alone.

\begin{figure}[!hbtp] \label{fig1}
\begin{center}
\includegraphics[width=5.3cm]{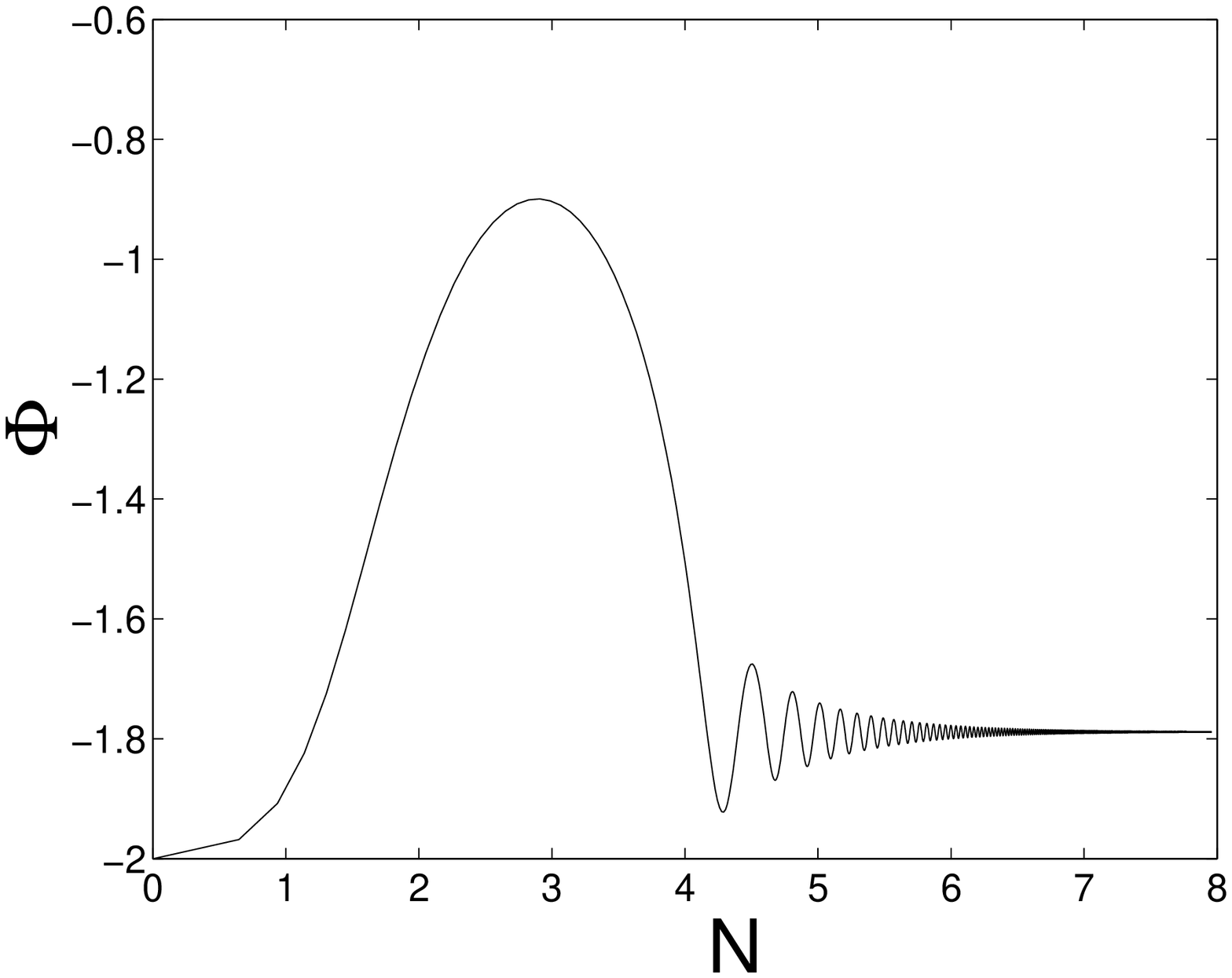}
\includegraphics[width=5.3cm]{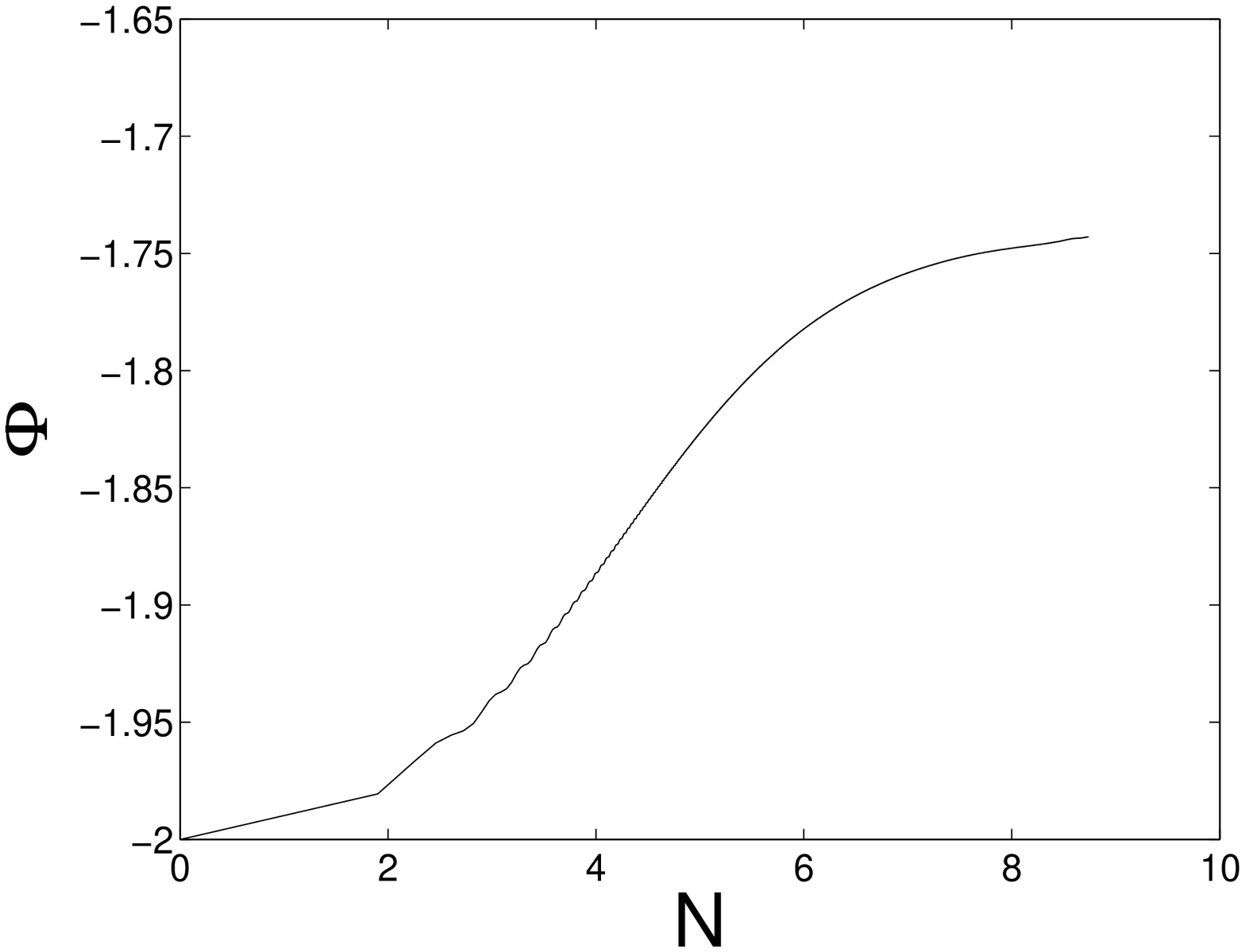}
\includegraphics[width=5.3cm]{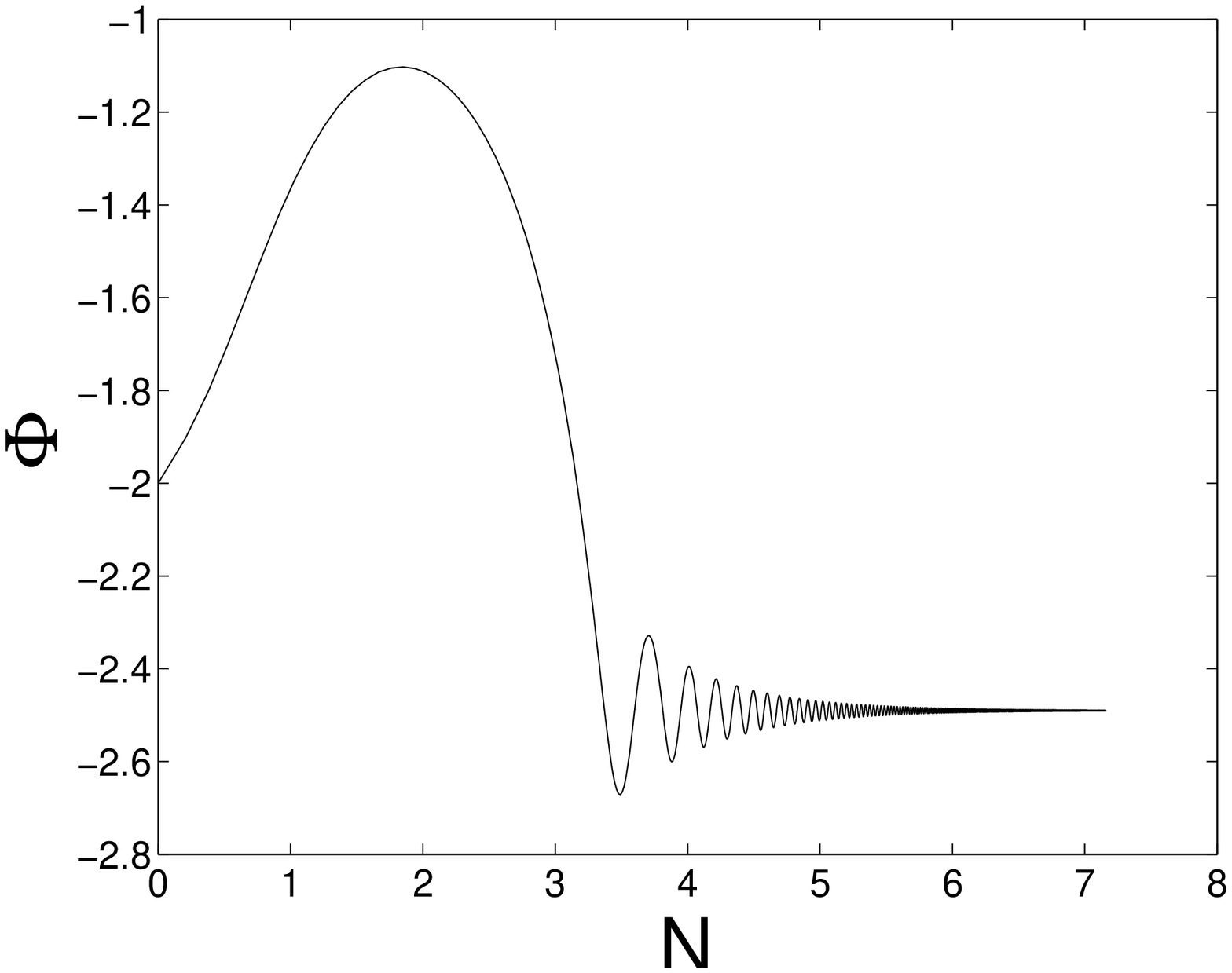}
\end{center}
\caption{Numerical solutions which shows the gravitational
potential $\Phi$ as a function of e-foldings after the end of
inflation $N$ for three types of initial conditions. Leftmost
panel: $\Phi=-2$, $\sigma_*=3$ and $\delta\sigma_*=0$. In this
case the $\sigma$-field dominates before it starts oscillating.
The final value of $\Phi$ agrees with eq. (\ref{Phibotox1}).
Middle panel: $\Phi=-2$, $\sigma_*=0.3$ and $\delta\sigma_*=0$. In
this case the $\sigma$-field does not dominate before it starts
oscillating. The final value of $\Phi$ also agrees with eq.
(\ref{Phibotox1}). Rightmost panel: $\Phi=-2$, $\sigma_*=3$ and
$\delta\sigma_*=1$. This panel is identical to fig. 3 of
\cite{Langlois:2004nn}.}
\end{figure}

\section{The Axion}

The main problem with the mechanism in the previous section
compared to the naive example in eq.(\ref{drh1}) is that the
inhomogeneities in the radiation fluid gravitationally sources
density perturbations in the $\sigma$-field fluid. In the case of
the periodic form of an axion field potential it is well known
that one can suppress the entropy perturbations in the axion
fluid. One might think that one can circumvent the problems using
this approach. However, as we will discuss below it will suffer
from similar difficulties.

Neglecting again the decay rates, the change in the curvature
perturbation $\zeta_{\sigma}$ in the $\sigma$-fluid is given by
its intrinsic non-adiabatic pressure perturbation
 \beq
\dot\zeta_{\sigma} = -\frac{H}{p_{\sigma}+\rho_{\sigma}}\delta
P_{intr}~,
 \eeq
where
 \beq
\delta P_{intr} = \delta
P_{\sigma}-c_{\sigma}^2\delta\rho_{\sigma}~.
 \eeq

In the ordinary curvaton limit, it is well known that
$\delta\sigma$ behaves like $\sigma$ and $\delta P_{intr}$
vanishes \cite{Sloth:2002xn}. It can be seen from
eq.(\ref{dsigma2}) that this holds even in the case where
$\delta\sigma_*=0$ because of the gravitational sourcing. If we
want $\dot\zeta_{\sigma}\neq 0$ in the relevant regimes, we need
that the perturbations evolve differently than the background.
This is true if the $\sigma$-field has a very non-linear periodic
potential and we find it useful to examine this more thoroughly.

Let us assume that the field $\sigma$ is an axionic field
for which the following potential is generated soon after the end
of inflation
 \beq
V(\sigma)
=\frac{1}{2}V_0\left(1-\cos\left(\frac{\sigma}{\sigma_0}\right)\right)~,
 \eeq
although one would have to worry, that a periodic potential will lead
to formation of topological defects \cite{Linde:1990yj}.

During the first ordinary slow-roll inflationary stage, the
potential of the $\sigma$-field and its derivatives are
insignificant, so the field fluctuation of the $\sigma$-field
generated during inflation is simply
 \beq
\delta\sigma^2=\frac{H_i^2}{4\pi^2}\ln\frac{aH_i}{k}~.
 \eeq
The linear approximation $\delta V(\sigma)\approx
(dV(\sigma)/d\sigma)\delta\sigma$ is only valid if the
$\sigma$-field and its dispersion is smaller than $\sigma_0$. As
mentioned above we are interested in the non-linear limit. To
understand the evolution of the density perturbations in this
limit we can adopt the strategy originally proposed by Kofman and
Linde \cite{Linde:1985yf,Kofman:1985zx,Kofman:1986wm}. At any
given scale $l=1/k$ the field will consist of a large scale
component that behaves as a homogeneous classical field
$\sigma_c(k)$ and a shortwave part $\delta\sigma$ corresponding to
momenta $k\geq 1/l$
 \beq \label{split1}
\sigma = \sigma_c(k)+\delta\sigma~.
 \eeq
The effective classical field at the scale $1/l$ is the sum of the
background vev $\tilde\sigma$ and its variance $\tilde\sigma_k$
 \beq
\sigma_c(k) =\sqrt{\tilde\sigma^2+\tilde\sigma_k^2}~.
 \eeq
For $\tilde\sigma_k>> \sigma_0$ this implies
 \beq
\sigma_c(k)^2 \approx \tilde\sigma_k^2 =
\frac{H_i^2}{4\pi^2}\ln\frac{k}{k_{min}}~.
 \eeq

It was shown in \cite{Linde:1985yf,Kofman:1985zx,Kofman:1986wm},
that the periodic nature of the axion field potential will then
efficiently power-law suppress the axion density perturbations
when $\sigma_0<< H$
 \beq
\frac{\delta\rho_{\sigma}}{\rho_{\sigma}}\simeq\frac{\rho_{\sigma}}{\rho}\frac{H}{2\pi\sigma_0}\cos\left(\tilde\sigma_c(k)/\sigma_0)\right)
\left(\frac{k}{aH} \right)^{\frac{H^2}{8\pi^2\sigma_0^2}}~.
 \eeq
while for a non-flat spectrum for $\delta\sigma$ the suppression
will be even exponential \cite{Enqvist:2001zp,Brustein:1998df}
 \beq \label{suppr}
\frac{\delta\rho_{\sigma}}{\rho_{\sigma}}\simeq\frac{\delta\sigma}{\sigma_0}
\exp\left(-\frac{1}{2}\int_k^{k_c}d\ln
k\frac{\delta\sigma^2}{\sigma_0^2} \right)
 \eeq
Above we used the approximations $\dot\sigma\approx 0$ and
$\delta\rho_{\sigma}\approx\delta V(\sigma)$.

However, while this can lead to a significant suppression of
entropy perturbations, in the separation of the field into the
classical field and its fluctuations, we have to take into account
that this separation will be different in each separate Hubble
volume at the end of inflation. The adiabatic density perturbations
can be understood to
be a consequence of inflation not ending simultaneously everywhere
due to the small inflaton fluctuations. At the end of inflation
the observable universe today consisted of a huge number of
causally separate Hubble patches whose relative evolution are
synchronized by a relative time delay due to the inflaton
fluctuations. No local physics can cancel this this
synchronization, but one can produce additional adiabatic
perturbations by converting entropy perturbation into adiabatic
perturbations in each Hubble volume separately. Because of this
time delay, the axion will also start to inflate and subsequently
oscillate at slightly different times synchronized to the inflaton
fluctuations in each different causally disconnected region. Thus,
the separation of the field into the classical field and its
fluctuations in eq.(\ref{split1}) will be shifted with respect to the
synchronized time delay between
each separate part of the universe, and if we take this into account the
initial adiabatic fluctuation will remain imprinted in the axion
fluid as it starts to dominate\footnote{I am grateful to Andrei
Linde, Slava Mukhanov and Misao Sasaki for this argument.}.

\overskrift{Acknowledgments}

\noindent I would like to acknowledge Andreas Albrecht, Andrei
Linde, Nemanja Kaloper, David Lyth, Slava Mukhanov, Misao Sasaki,
Lorenzo Sorbo and David Wands for comments and criticism.
Especially I would like to thank Andrei Linde, David Lyth, Slava Mukhanov,
Misao Sasaki and David Wands for pointing out a problem in the first
version. The work was supported in part by the DOE Grant DE-FG03-91ER40674.

\end{document}